\documentclass[seceq]{ptptex}

\notypesetlogo                       

\title{Properties of the Ground-State $q \bar{q}$ Mesons 
 and \\ Possible Classification of Observed Mesons in the
 $\widetilde{U}(12)_{SF} \times O(3,1)_{L}$ Scheme}

\author{Kenji \textsc{\scshape Yamada}%
}

\inst{Department of Engineering Science,
 Junior College Funabashi Campus, \\
 Nihon University, Funabashi 274-8501, Japan
}

\abst{We examine possible assignments for known
 mesons to the predicted ground-state $q \bar{q}$
 multiplets in the $\widetilde{U}(12)_{SF}$-classification scheme
 of hadrons, based on their observed properties.
 The masses of missing members of the ground-state multiplets
 are estimated, using a phenomenological mixing scheme for the normal-
 and extra-spin wave functions specific to this classification scheme.
 We also examine experimental candidates for the excited
 $1P$, $1D$ and $2S$ multiplets in a framework of the
 $\widetilde{U}(12)_{SF} \times O(3,1)_{L}$-classification scheme,
 based on the analysis of the ground-state masses.
 Then we see that the two exotic mesons
 $\pi _{1}(1400)$ and $\pi _{1}(1600)$ with $J^{PC}=1^{-+}$
 have suitable masses to be assigned to the $P$-wave
 excitation and also there are a number of observed mesonic states which
 could be classified in terms of the
 $\widetilde{U}(12)_{SF} \times O(3,1)_{L}$ scheme.}

\begin{document}

\maketitle

\section{Introduction
}
Recently, Ishida et al. have proposed the
 covariant $\widetilde{U}(12)_{SF}$-classification scheme of
 hadrons with $\widetilde{U}(12)_{SF} \times O(3,1)_{L}$,\cite{IIM2000}
 which gives covariant quark representations for
 composite hadrons with definite Lorentz and chiral transformation
 properties. The $\widetilde{U}(12)_{SF}$-classification scheme
 has a ``static'' unitary $U(12)_{SF}$ spin-flavor symmetry in the rest frame of
 hadrons,\cite{IIYMO2005} embedded in the covariant
 $\widetilde{U}(12)_{SF}$-representation space, which includes
 subgroups as
 $\widetilde{U}(12)_{SF} \supset \widetilde{U}(4)_{D} \times U(3)_{F}$
 ($\widetilde{U}(4)_{D}$ being the pseudounitary homogeneous
 Lorentz group for Dirac spinors). Since
\begin{subequations}
	\begin{equation}
		U(12)_{SF} \supset U(4)_{D} \times U(3)_{F}
	\end{equation}
with
	\begin{equation}
		U(4)_{D} \supset SU(2)_{\rho} \times SU(2)_{\sigma},
		\label{eq:1.1b}
	\end{equation}
\end{subequations}
 the static $U(12)_{SF}$ symmetry includes as its subgroup
 both the nonrelativistic spin-flavor $SU(6)_{SF}$ and the chiral
 $U(3)_{L} \times U(3)_{R}$ symmetry as
\begin{subequations}
	\begin{equation}
		U(12)_{SF} \supset SU(6)_{SF} \times SU(2)_{\rho}
	\end{equation}
and
	\begin{equation}
		U(12)_{SF} \supset U(3)_{L} \times U(3)_{R} \times SU(2)_{\sigma},
	\end{equation}
\end{subequations}
 where $SU(2)_{\rho}$ and $SU(2)_{\sigma}$ are the Pauli-spin
 groups concerning the boosting and intrinsic spin rotation,
 respectively, of constituent quarks (being connected with decomposition
 of Dirac $\gamma$-matrices, $\gamma \equiv \rho \otimes \sigma$).
 This implies that the $\widetilde{U}(12)_{SF}$-classification scheme
 is able to incorporate effectively the effects of chiral symmetry
 and its spontaneous breaking, essential for understanding
 of properties of the low-lying hadrons, into what is called
 a constituent quark model.

\section{Experimental candidates for the ground-state $q \bar{q}$ mesons
}

\subsection{Essential features of the
 $\widetilde{U}(12)_{SF}$-classification scheme
}
An essential feature of the $\widetilde{U}(12)_{SF}$-classification
 scheme is to have the static
 $U(4)_{D}$-spin symmetry in Eq. (\ref{eq:1.1b}) for light $u, d, s$
 quarks confined inside hadrons.
 The degree of freedom on the $\rho$-spin, being indispensable
 for covariant description of spin $1/2$ particles,
 offers a basis to define the rule of chiral transformation
 for quark-composite hadrons.\cite{IIM2000, IIYMO2005}
 
Since we have the $\rho$-spin degree of freedom, which is
 discriminated by the eigenvalues $r=\pm$ of $\rho _{3}$,
 in addition to the ordinary Pauli-spin, the ground states
 of light-quark $q\bar{q}$ mesons are composed of eight
 $SU(3)_{F}$ multiplets with respective $J^{PC}$ quantum numbers,
 two pseudoscalars
 $\{P^{(N)}(0^{-+}), P^{(E)}(0^{-+})\}$, two scalars
 $\{S_{A}^{(N)}(0^{++}), S_{B}^{(E)}(0^{+-})\}$, two vectors
 $\{V^{(N)}(1^{--}), V^{(E)}(1^{--})\}$, and two axial-vectors
 $\{A^{(N)}(1^{++}), B^{(E)}(1^{+-})\}$
 ($N$ and $E$ denoting ``normal'' and ``extra''),
 where each $N (E)$ even-parity multiplet is the chiral partner
 of the corresponding $N (E)$ odd-parity multiplet and
 they form linear representations of the
 chiral $U(3)_{L} \times U(3)_{R}$ symmetry.
 
For heavy-light mesons we have two heavy-spin multiplets
 $\{P(0^{-}), V(1^{-})\}$ and $\{S(0^{+}), A(1^{+})\}$, 
 which are the chiral partner of
 each other, since the eigenstates only with the $\rho _{3}$-eigenvalue
 of $r=+$ are taken for heavy quarks. For heavy-heavy mesons we have
 the same $\{P(0^{-}), V(1^{-})\}$-spin multiplets as in the conventional
 quark model. For both the heavy-light and heavy-heavy systems, their spin
 wave functions are the perfect mixtures, equally weighted sum,
 of the $N$ and $E$ states concerning the respective $J^{P}$ states.

\subsection{Experimental candidates for the ground-state $q\bar{q}$
 mesons in the
 $\widetilde{U}(12)_{SF}$-\hspace{0pt}classification scheme}
\label{expcand}
We try to assign some of the known mesons to the predicted
 ground-state $q\bar{q}$ multiplets, resorting to their $J^{PC}$
 quantum numbers and masses. The experimental data are taken
 from the Particle Data Group 2004 edition.\cite{PDG2004}
 The resulting assignments, though some of them are ambiguous,
 are shown in the Table \ref{table:1}.

\begin{table}
  \caption{Experimental candidates for the ground-state $q\bar{q}$
  mesons in the $\widetilde{U}_{SF}(12)$-classification scheme. The data
  are taken from Ref.~\citen{PDG2004}.}
  \label{table:1}
  \begin{center}
  \renewcommand{\arraystretch}{1.3}
    \begin{tabular}{ccccccccc} \hline \hline
	& $P^{(N)}$ & $S_{A}^{(N)}$ & $P^{(E)}$ & $S_{B}^{(E)}$ & $V^{(N)}$ & $A^{(N)}$ & $V^{(E)}$ & $B^{(E)}$ \\ \cline{2-9}
	$q\bar{q}$ & \multicolumn{4}{c}{$1^{1}S_{0}$} & \multicolumn{4}{c}{$1^{3}S_{1}$} \\
		& $0^{-+}$ & $0^{++}$ & $0^{-+}$ & $0^{+-}$ & $1^{--}$ & $1^{++}$ & $1^{--}$ & $1^{+-}$ \\ \hline
	$n\bar{n}$ & $\pi$ & $a_{0}(980)$ & $\pi(1300)$ &  & $\rho(770)$ & $a_{1}(1260)$ &   & $b_{1}(1235)$ \\
		& $\eta$ & $\sigma$ & $\eta(1295)$ &   & $\omega(782)$ & $f_{1}(1285)$ &   & $h_{1}(1170)$ \\
	$s\bar{s}$ & $\eta '(958)$ & $f_{0}(980)$ & $\eta(1475)$ &   & $\phi(1020)$ & $f_{1}(1420)$ &   & $h_{1}(1380)$ \\
	$s\bar{n}$ & $K$ & $\kappa$ & $K(1460)$ &   & $K^{*}(892)$ & $K_{1}(1270)$ & $K^{*}(1410)$ & $K_{1}(1400)$ \\
	$c\bar{n}$ & $D$ &   & -- & -- & $D^{*}$ &   & -- & -- \\
	$c\bar{s}$ & $D_{s}$ & $D^{*}_{sJ}(2317)$ & -- & -- & $D^{*}_{s}$ & $D_{sJ}(2460)$ & -- & -- \\
	$b\bar{n}$ & $B$ &   & -- & -- & $B^{*}$ &   & -- & -- \\
	$b\bar{s}$ & $B_{s}$ &   & -- & -- & $B^{*}_{s}$ &   & -- & -- \\
	$c\bar{c}$ & $\eta _{c}(1S)$ & -- & -- & -- & $J/\psi(1S)$ & -- & -- & -- \\
	$b\bar{b}$ & $\eta _{b}(1S)$ & -- & -- & -- & $\Upsilon(1S)$ & -- & -- & -- \\ \hline
	\end{tabular}
  \end{center}
\end{table}

Here we make some comments on these assignments as follows:

\begin{enumerate}
\renewcommand{\labelenumi}{(\roman{enumi})}
	\item The light scalar mesons $\{a_{0}(980), \sigma, f_{0}(980),
	\kappa\}$ are assigned to the $S_{A}^{(N)}(0^{++})$ nonet
	as a chiral partner of the
	$\pi$-meson $P^{(N)}(0^{-+})$ nonet. Recently the existence of
	the $\kappa$ meson has been confirmed in two independent
	partial-wave analyses of the decay
	$J/\psi \rightarrow \bar{K}^{*}(892)^{0}K^{+}\pi^{-}$
	by the BES Collaboration.\cite{BES}
	\item The $B^{(E)}(1^{+-})$ nonet is composed of the
	$b_{1}(1235)$, $h_{1}(1170)$, $h_{1}(1380)$ and $K_{1}(1400)$
	mesons.
	The low-mass vector meson $K^{*}(1410)$ is assigned as
	a member of the $V^{(E)}(1^{--})$ nonet which is a chiral partner
	of the $B^{(E)}(1^{+-})$ nonet.
	\item The axial-vector mesons $\{a_{1}(1260), f_{1}(1285),
	f_{1}(1420), K_{1}(1270)\}$ are assigned to the
	$A^{(N)}(1^{++})$ nonet which is a chiral partner
	of the $\rho(770)$-meson $V^{(N)}(1^{--})$ nonet.
	The $a_{1}(1260)$ and $f_{1}(1285)$ mesons are tentatively
	assigned to this nonet, while their masses seem to be
	higher than could be expected and also their observed properties
	of radiative transitions does not seem to be consistent
	with those expected in the $\widetilde{U}(12)_{SF}$-classification
	scheme.\cite{Maeda}
	\item The recent discovered mesons
	$D_{sJ}^{*}(2317)$ and $D_{sJ}(2460)$ are just assigned to the
	$\{S(0^{+}), A(1^{+})\}$-spin multiplet as a chiral partner of the
	$\{P(0^{-}), V(1^{-})\}$-spin multiplet,
	$\{D_{s}, D_{s}^{*}\}$.\cite{Ishida2003} \ These newly
	observed states, together with the $\sigma$-meson nonet,
	are the best candidates for the hadronic states with
	the $\rho _{3}$-eigenvalue of $r=-$ whose existence is expected in the
	$\widetilde{U}(12)_{SF}$-classification scheme.
	\item The $N$ and $E$ states with
	the same $J^{PC}$, that is, the vector $V^{(N,E)}$
	and pseudoscalar $P^{(N,E)}$ states generally mix together,
	due to the spontaneous as well as explicit breaking
	of chiral symmetry, and so do the strange scalar
	$\kappa^{(N,E)}$ and axial-vector $K_{1}^{(N,E)}$ states.
	However, as far as the pseudoscalar $P^{(N,E)}$ octet is concerned,
	it is assumed that no mixing occurs so as to preserve the property of
	both the $\pi$-meson octet, the Nambu-Goldstone bosons
	associated with the spontaneous breaking of the axial
	$SU(3)_{A}$ symmetry, and the $\sigma$-meson nonet belonging
	to the same chiral multiplet.
	As for the flavor-singlet $P^{(N)}$ and $P^{(E)}$ states,
	the $\eta'$-like mesons,
	there would be mixing between them according to
	a common understanding that the axial $U(1)_{A}$ is
	not a true symmetry in the strong interactions.
\end{enumerate}

\section{Masses and $N$-$E$ mixings of the ground-state
 $q \bar{q}$ mesons}
\label{masses}
In the following we examine the respective $N$- and
 $E$-state mixings of
 $V^{(N,E)}$, $K_{1}^{(N,E)}$ and $\kappa^{(N,E)}$
 in a simple phenomenological model.
 Through this analyses the flavor mixing between $n \bar{n}$
 and $s \bar{s}$ states is neglected
 for the isoscalar channels and the masses of $n \bar{n}(I=0)$ states
 are taken to be equal to those of $n \bar{n}(I=1)$ states.
 It is also assumed that the $N$- and $E$-state masses of
 different flavor states are related to each other as
\begin{equation}
	M_{(N,E)}(s \bar{s}) - M_{(N,E)}(s \bar{n}) =
	M_{(N,E)}(s \bar{n}) - M_{(N,E)}(n \bar{n})
	\equiv \Delta m_{s},
	\label{eq:deltams}
\end{equation}
where $\Delta m_{s}$ is taken to be 120 MeV.

\subsection{Phenomenological mixing scheme of
 the N and E states
}
In the normal and extra basis
 $\left( |E \rangle, |N \rangle \right)$,
 the mass-squared matrix describing the $N$-$E$ mixing
 can be written as
\begin{equation}
	\boldsymbol{M}^{2} =
		\begin{pmatrix}
			M_{E}^{2} & \Delta \\
			\Delta & M_{N}^{2}
		\end{pmatrix},
\end{equation}
where $M_{N}$ and $M_{E}$ are the masses of the $N$ and
 $E$ states, respectively, before mixing and $\Delta$
 is a phenomenological parameter corresponding to the mixing
 strength. Assuming that the physical-state basis
 $\left( |L \rangle, |H \rangle \right)$ ($L$ and $H$
 denoting ``low'' and ``high'') is an eigenvector of the
 mass-squared matrix $\boldsymbol{M}^{2}$ with the
 eigenvalues $M_{L}^{2}$ and $M_{H}^{2}$, we can diagonalize
 the matrix $\boldsymbol{M}^{2}$ as
\begin{equation}
	U^{-1}\boldsymbol{M}^{2}U =
		\begin{pmatrix}
			M_{L}^{2} & 0 \\
			0 & M_{H}^{2}
		\end{pmatrix}
\end{equation}
 by the unitary transformation
\begin{subequations}
	\begin{equation}
		\begin{pmatrix}
			|L \rangle \\ |H \rangle
		\end{pmatrix}
		= U^{-1}
		\begin{pmatrix}
			|E \rangle \\ |N \rangle
		\end{pmatrix}
	\end{equation}
 with
	\begin{equation}
		U^{-1} =
			\begin{pmatrix}
				\cos \theta & \sin \theta \\
				-\sin \theta & \cos \theta
			\end{pmatrix},
	\end{equation}
\end{subequations}
in which the $N$-$E$ mixing angle is defined.
 It is noted that the low- and high-mass states have
 dominantly $|E \rangle$ and $|N \rangle$ components,
 respectively, if the mixing angle is $|\theta| < 45^{\circ}$.

\subsection{Masses and mixing properties of
 the vector $V^{(N)}$ and $V^{(E)}$ nonets}
\label{vector}
In the analysis of the vector meson nonets $V^{(N)}(1^{--})$
 and $V^{(E)}(1^{--})$ we assume that
 the $V^{(N)}$ and $V^{(E)}$ states are mixed maximally,
 that is, the $N$-$E$ mixing angle is $|\theta| = 45^{\circ}$,
 in accord with the fact that the lowest-lying vector mesons
 $\rho(770)$, $\omega(782)$, $\phi(1020)$, $K^{*}(892)$
 are well described as nonrelativistic $q\bar{q}$ states and
 the analysis of their radiative transitions
 in the $\widetilde{U}(12)_{SF}$-classification scheme.\cite{Maeda}
 These maximally mixed low- and high-mass vector states,
 $V^{(NR)}$ and $V^{(ER)}$, mean to be nonrelativistic (NR) and
 extremely relativistic (ER) states which have the $\rho_{3}$-eigenvalues
 of $(r_{q}, r_{\bar q}) = (+, +)$ and $(-, -)$, respectively.

First, we consider the $K^{*}$ system, whose mass-squared
 matrix relation is given by
\begin{equation}
	U^{-1}
	\begin{pmatrix}
		M_{K^{*(E)}}^{2} & \Delta_{K^{*}} \\
		\Delta _{K^{*}} & M_{K^{*(N)}}^{2}
		\end{pmatrix}U =
		\begin{pmatrix}
			M_{K^{*}(892)}^{2} & 0 \\
			0 & M_{K^{*}(1410)}^{2}
		\end{pmatrix},
\end{equation}
where $K^{*}(892)$ and $K^{*}(1410)$ are members of the 
 $V^{(N)}(1^{--})$ and $V^{(E)}(1^{--})$ nonets
 as mentioned in \S\ref{expcand}.
 Using the mixing angle
 $|\theta| = 45^{\circ}$, which means
 $M_{K^{*(N)}} = M_{K^{*(E)}}$, and $M_{K^{*}(892)} = 894\ \mathrm{MeV}$
 and $M_{K^{*}(1410)} = 1414\ \mathrm{MeV}$,
 we obtain
\begin{equation}
	M_{K^{*(N,E)}} = 1183\ \mathrm{MeV}, \ \ 
	|\Delta _{K^{*}}| = 0.6001\ \mathrm{GeV}^{2}.
\end{equation}
 Then, from the assumption in Eq. (\ref{eq:deltams}) we have
 $M_{\rho^{(N,E)}} = M_{\omega^{(N,E)}} = 1063\ \mathrm{MeV}$
 for the $\rho$ and $\omega$ systems. For the $\rho$ system
 the mass-squared matrix relation is given by
\begin{equation}
	U^{-1}
		\begin{pmatrix}
			M_{\rho^{(E)}(1063)}^{2} & \Delta _{\rho} \\
			\Delta_{\rho} & M_{\rho^{(N)}(1063)}^{2}
		\end{pmatrix}U =
		\begin{pmatrix}
			M_{\rho(770)}^{2} & 0 \\
			0 & M_{\rho^{\prime}}^{2}
		\end{pmatrix},
\end{equation}
 which gives,
 taking the mass value of
 $M_{\rho(770)} = M_{\omega(782)} = 776\ \mathrm{MeV}$,
\begin{equation}
	M_{\rho^{\prime}} = M_{\omega^{\prime}} = 1290\ \mathrm{MeV}, \ \ 
	|\Delta _{\rho}| = |\Delta _{\omega}| = 0.5284\ \mathrm{GeV}^{2}.
\end{equation}
 In a similar way, we obtain
\begin{subequations}
	\begin{equation}
		M_{\phi^{(N,E)}} = 1303\ \mathrm{MeV}, \ \
		M_{\phi^{\prime}} = 1535\ \mathrm{MeV},
	\end{equation}
	\begin{equation}
		|\Delta _{\phi}| = 0.6585\ \mathrm{GeV}^{2}
	\end{equation}
\end{subequations}
 for the $\phi$ system.
 These results are tabulated in Table \ref{table:2}.
 
 Here we point out that there exist observed candidates
 suitable for the predicted low-mass vector mesons $\rho'(1290)$
 and $\omega'(1290)$. In fact, there is some experimental
 evidence for the $\rho(1250)$ reported by the LASS\cite{LASS}
 and OBELIX\cite{OBELIX} Collaborations and
 the existence of $\omega(1250)$ is claimed in the analysis
 of the $e^{+}e^{-} \to \pi^{+}\pi^{-}\pi^{0}$ cross section
 by the SND\cite{SND} and BABAR\cite{BABAR} Collaborations.
 Furthermore, a recent reanalysis of the BABAR\cite{BABAR},
 SND\cite{SND,SNDop} and CMD-2\cite{CMD2} data
 on the $e^{+}e^{-} \to \pi^{+}\pi^{-}\pi^{0}$ and
 $e^{+}e^{-} \to \omega\pi^{0}$ cross sections indicates
 the existence of these two vector states.\cite{Komada}

It is worthwhile to mention that only the $V^{(N)}(1^{--})$
 states and not the $V^{(E)}(1^{--})$ can be produced
 in the $e^{+}e^{-}$ annihilation
 process of $e^{+}e^{-} \to \gamma^{*} \to V$ due to the
 chirality conservation of quarks,\footnote{A vertex operator
 for the quark-photon interaction is assumed to be proportional to
 $\gamma_{\mu}$.} since both the constituent
 quark and antiquark of $V^{(N)}(1^{--})$ have the same
 chirality, while those of $V^{(E)}(1^{--})$ have opposite
 one in the $\widetilde{U}(12)_{SF}$-classification scheme.
 Therefore we can expect that the predicted $\rho'(1290)$ and
 $\omega'(1290)$ mesons, which have the same amount of
 $V^{(N)}(1^{--})$ component as the $\rho(770)$ and
 $\omega(782)$, are seen in the above
 $e^{+}e^{-}$ annihilation process.

\begin{table}
  \caption{Masses and mixing properties of the vector meson
	$V^{(N)}(1^{--})$ and $V^{(E)}(1^{--})$ nonets. Fitted values
	are underlined.}
  \label{table:2}
  \begin{center}
  \renewcommand{\arraystretch}{1.3}
    \begin{tabular}{cccccc} \hline \hline
	&   & Predicted mass &  Observed mass$^{\textrm{a)}}$ &  Mixing angle $|\theta|$ & Mixing strength $|\Delta|$ \\
	$q\bar{q}$ & State & (MeV) & (MeV) &   & $(\mathrm{GeV}^{2})$ \\ \hline
	$n\bar{n}$ & $\rho(770)$ & $\underline{776}$ & $775.8\pm 0.5$ & $45^{\circ}$ & 0.5284 \\
		& $\rho'$ & 1290 & -- &   \\ \cline{2-4}
		& $\omega(782)$ & 776 & $782.59\pm 0.11$ \\
		& $\omega'$ & 1290 & -- &   \\ \hline
	$s\bar{s}$ & $\phi(1020)$ & $\underline{1019}$ & $1019.456\pm 0.020$ & $45^{\circ}$ & 0.6585 \\
		& $\phi'$ & 1535 & -- &   \\ \hline
	$s\bar{n}$ & $K^{*}(892)$ & $\underline{894}$ & $893.9\pm 2.5$ & $45^{\circ}$ & 0.6001 \\
		& $K^{*}(1410)$ & $\underline{1414}$ & $1414\pm 15$   \\ \hline
	\multicolumn{6}{l}{a) The data are taken from Ref.~\citen{PDG2004}.} \\
	\end{tabular}
  \end{center}
\end{table}

\subsection{Masses and mixing properties of
 the axial-vector $A^{(N)}$ and $B^{(E)}$ nonets
}
In the axial-vector meson nonets, $A^{(N)}(1^{++})$ and
 $B^{(E)}(1^{+-})$, their mixing could occur only for
 the strange $K_{1}$ system and therefore the $N$ and $E$ states
 are physical ones for the isoscalar and isovector channels.
 In this analysis we select, as input data,
 the $b_{1}(1235)$, $K_{1}(1270)$ and $K_{1}(1400)$ mesons
 from among the assigned mesons in \S\ref{expcand}.

For the $K_{1}$ system we obtain
 $M_{K_{1B}^{(E)}} = 1350\ \mathrm{MeV}$
 from the assumption in Eq. (\ref{eq:deltams}) with
 the measured mass of $M_{b_{1}(1235)} = 1230\ \mathrm{MeV}$
 and the mass-squared matrix
 relation is given by
\begin{equation}
	U^{-1}
		\begin{pmatrix}
		M_{K_{1B}^{(E)}(1350)}^{2} & \Delta _{K_{1}} \\
		\Delta _{K_{1}} & M_{K_{1A}^{(N)}}^{2}
		\end{pmatrix}U =
		\begin{pmatrix}
			M_{K_{1}(1270)}^{2} & 0 \\
			0 & M_{K_{1}(1400)}^{2}
		\end{pmatrix}.
\end{equation}
Taking the mass values of $M_{K_{1}(1270)} = 1273\ \mathrm{MeV}$
 and $M_{K_{1}(1400)} = 1402\ \mathrm{MeV}$, we derive
\begin{subequations}
	\begin{equation}
		M_{K_{1A}^{(N)}} = 1328\ \mathrm{MeV},
	\end{equation}
	\begin{equation}
		|\Delta _{K_{1}}| = 0.1700\ \mathrm{GeV}^{2}, \ \
		|\theta _{K_{1}}| = 49.9^{\circ}
	\end{equation}
\end{subequations}
and then
\begin{subequations}
	\begin{align}
		M_{a_{1}} &= M_{f_{1}(n\bar{n})} = 1210\ \mathrm{MeV}, \ \
		M_{f_{1}'(s\bar{s})} = 1450\ \mathrm{MeV}, \\
		M_{b_{1}} &= M_{h_{1}(n\bar{n})} = 1230\ \mathrm{MeV}, \ \
		M_{h_{1}'(s\bar{s})} = 1470\ \mathrm{MeV}
		\label{eq:3.13b}
	\end{align}
\end{subequations}
for the isoscalar and isovector channels.
These results are given in Table \ref{table:3}.

It should be noted here that $K_{1A}^{(N)}$ and $K_{1B}^{(E)}$
 are the $1^{3}P_{1}$ and $1^{1}P_{1}$ states respectively
 in the conventional quark model, while they are relativistic
 $S$-wave states in which each $q$ and $\bar{q}$ has
 the opposite $\rho _{3}$-eigenvalue in the
 $\widetilde{U}(12)_{SF}$-classification scheme.
 The resulting mixing angle $|\theta _{K_{1}}| = 49.9^{\circ}$
 means that the dominant components of $K_{1}(1270)$ and
 $K_{1}(1400)$ are the $K_{1A}^{(N)}$ and $K_{1B}^{(E)}$ states,
 respectively.

\begin{table}
 \caption{Masses and mixing properties of the axial-vector
	meson $A^{(N)}(1^{++})$ and $B^{(E)}(1^{+-})$ nonets.
	Fitted values are underlined.}
 \label{table:3}
 \begin{center}
 \renewcommand{\arraystretch}{1.3}
 	\begin{tabular}{cccccc} \hline \hline
	&   & Predicted mass & Observed mass$^{\textrm{a)}}$ & Mixing angle $|\theta|$ & Mixing strength $|\Delta|$ \\
	$q\bar{q}$ & State & (MeV) & (MeV) &   & $(\mathrm{GeV}^{2})$ \\ \hline
	$n\bar{n}$ & $a_{1}(1260)$ & 1210 & $1230\pm 40$ & -- & -- \\
		& $f_{1}(1285)$ & 1210 & $1281.8\pm 0.6$ & -- & -- \\ \cline{2-4}
		& $b_{1}(1235)$ & $\underline{1230}$ & $1229.5\pm 3.2$ & -- &-- \\
		& $h_{1}(1170)$ & 1230 & $1170\pm 20$ & -- & -- \\ \hline
	$s\bar{s}$ & $f_{1}(1420)$ & 1450 & $1426.3\pm 0.9$ & -- & -- \\
		& $h_{1}(1380)$ & 1470 & $1386\pm 19$ & -- & -- \\ \hline
	$s\bar{n}$ & $K_{1}(1270)$ & $\underline{1273}$ & $1273\pm 7$ & $49.9^{\circ}$ & 0.1700 \\
		& $K_{1}(1400)$ & $\underline{1402}$ & $1402\pm 7$ \\ \hline
	\multicolumn{6}{l}{a) The data are taken from Ref.~\citen{PDG2004}.} \\
	\end{tabular}
 \end{center}
\end{table}

\subsection{Masses and mixing properties of
 the scalar $S_{A}^{(N)}$ and $S_{B}^{(E)}$ nonets
}
The scalar $S_{A}^{(N)}(0^{++})$ and $S_{B}^{(E)}(0^{+-})$
 states could mix only for the strange $\kappa$ system,
 as in the $K_{1}$ system. For lack of experimental information
 and theoretical understanding on the light scalar mesons,
 we assume here that the mixing strength between $\kappa _{A}^{(N)}$
 and $\kappa _{B}^{(E)}$ is equal to that of the $K_{1}$
 system, that is,
\begin{equation}
	\Delta _{\kappa} = \Delta _{K_{1}},
\end{equation}
which might be realized, provided that the
 $\kappa _{A}^{(N)}$-$\kappa _{B}^{(E)}$ mixing originates
 from the spontaneous breaking of chiral symmetry.
 As input data in this analysis we take the mass values of
 the $a_{0}(980)$ and $\kappa$ mesons to be
 $985\ \mathrm{MeV}$ and $875\ \mathrm{MeV}$, respectively.

For the $\kappa$ system we have
 $M_{\kappa _{A}^{(N)}} = 1105\ \mathrm{MeV}$
 from the assumption in Eq. (\ref{eq:deltams}) with
 the $a_{0}(980)$ mass and
 the mass-squared matrix relation is given by
\begin{equation}
	U^{-1}
		\begin{pmatrix}
		M_{\kappa _{B}^{(E)}}^{2} & \Delta _{\kappa} \\
		\Delta _{\kappa} & M_{\kappa _{A}^{(N)}(1105)}^{2}
		\end{pmatrix}U =
		\begin{pmatrix}
			M_{\kappa(875)}^{2} & 0 \\
			0 & M_{\kappa'}^{2}
		\end{pmatrix}
\end{equation}
 with $|\Delta _{\kappa}| = 0.1700\ \mathrm{GeV}^{2}$.
 This relation gives
\begin{subequations}
	\begin{equation}
		M_{\kappa _{B}^{(E)}} = 911\ \mathrm{MeV}, \ \
		M_{\kappa'} = 1135\ \mathrm{MeV},
	\end{equation}
	\begin{equation}
		|\theta _{\kappa}| = 20.5^{\circ}
	\end{equation}
\end{subequations}
and then
\begin{subequations}
	\begin{align}
		M_{a_{0}} &= M_{f_{0}(n\bar{n})} = 985\ \mathrm{MeV}, \ \
		M_{f_{0}'(s\bar{s})} = 1225\ \mathrm{MeV}, \\
		M_{b_{0}} &= M_{h_{0}(n\bar{n})} = 790\ \mathrm{MeV}, \ \
		M_{h_{0}'(s\bar{s})} = 1030\ \mathrm{MeV}
		\label{eq:3.17b}
	\end{align}
\end{subequations}
for the isoscalar and isovector channels.
 These results are given in Table \ref{table:4}.

 Here it is noticeable that a dominant component of the
 $\kappa(875)$ is $\kappa _{B}^{(E)}$ while that of the $\kappa'(1135)$
 is $\kappa _{A}^{(N)}$, due to the mixing angle of
 $|\theta _{\kappa}| = 20.5^{\circ}$. This implies that
 the unknown $\kappa'(1135)$ rather than the $\kappa(875)$ is
 a member of the light scalar $\sigma$-meson nonet.
 In this connection, it is of great interest that
 a recent lattice-QCD study
 of light scalar mesons with the interpolating field
 $\bar{\psi}\psi$ gives
 the results of the lightest $a_{0}$ meson having a mass of
 $1.01\pm 0.04\ \mathrm{GeV}$ and the $K_{0}^{*}$ meson
 $100$-$130\ \mathrm{MeV}$ heavier than the $a_{0}$ meson.\cite{UKQCD}
 These results are in conformity with our prediction,
 since the $\kappa _{A}^{(N)}$ and $\kappa _{B}^{(E)}$ are considered
 to be states corresponding to the interpolating fields\footnote{
 Chiral transformation properties of various interpolating fields
 of mesons have been examined in Ref.~\citen{Glozman}.}
 $\bar{\psi}\psi$ and $\bar{\psi}i\gamma _{\mu}\partial_{\mu}\psi$,
 respectively.

\begin{table}
 \caption{Masses and mixing properties of the scalar meson
	$S^{(N)}_{A}(0^{++})$ and $S^{(E)}_{B}(0^{+-})$ nonets.
	Fitted values are underlined.}
 \label{table:4}
 \begin{center}
 \renewcommand{\arraystretch}{1.3}
	\begin{tabular}{cccccc} \hline \hline
	&   & Predicted mass & Observed mass$^{\textrm{a)}}$ 
	&  Mixing angle $|\theta|$ & Mixing strength $|\Delta|$ \\
	$q\bar{q}$ & State & (MeV) & (MeV) &   & $(\mathrm{GeV}^{2})$ \\ \hline
	$n\bar{n}$ & $a_{0}(980)$ & $\underline{985}$ & $ 984.7\pm 1.2$ & -- & -- \\
		& $\sigma$ & 985 & $\sim$ 400-600 & -- & -- \\ \cline{2-4}
		& $b_{0}$ & 790 &   & -- & -- \\
		& $h_{0}$ & 790 &   & -- & -- \\ \hline
	$s\bar{s}$ & $f_{0}(980)$  & 1225 & $ 980\pm 10$ & -- & -- \\
		& $h_{0}'$ & 1030 &   & -- & -- \\ \hline
	$s\bar{n}$ & $\kappa'$ & 1135 &   & $20.5^{\circ}$ & 0.1700 \\
		& $\kappa$ & $\underline{875}$ & $878\pm 23^{+64}_{-55}$ &  \\ \hline
	\multicolumn{6}{l}{a) The data are taken from Ref.~\citen{PDG2004}, unless the mass of the $\kappa$ from Ref.~\citen{BES}.} 
	\end{tabular}
  \end{center}
\end{table}

\section{Mass spectra and possible assignments
 for observed mesons in the
 $\widetilde{U}(12)_{SF} \times O(3,1)_{L}$-classification scheme}
 
\subsection{Mass spectra of $n\bar{n}$ mesons}

We examine excited states of the respective ground-state sectors
 in the extended $\widetilde{U}(12)_{SF}$-classification scheme with
 $\widetilde{U}(12)_{SF} \times O(3,1)_{L}$, in which the degree of
 freedom concerning the orbital motion of quarks is incorporated.
 In this classification scheme it is predicted that there exist
 some $q\bar{q}$
 exotic states with $J^{PC}$ quantum numbers, such as $0^{+-}$
 in the ground states,
 $0^{--}$ and $1^{-+}$ in the excited $P$-wave states,
 $2^{+-}$ in the excited $D$-wave states, which never appear
 in any nonrelativistic quark model.
 There is presently the experimental observation of two
 exotic mesons with $J^{PC} = 1^{-+}$,
 the $\pi _{1}(1400)$ and $\pi _{1}(1600)$.\cite{PDG2004}
 Since, as far as the $1^{-+}$ exotic state
 is concerned, we have just two states in the $P$-wave excitation
 in this scheme,
 the $\pi _{1}(1400)$ and $\pi _{1}(1600)$ mesons are
 expected to be promising experimental candidates
 for these predicted exotics.
 
In the $\widetilde{U}(12)_{SF} \times O(3,1)_{L}$-classification scheme
 the mass of excited states is given by
\begin{equation}
\label{eq:4.1}
	M_{N}^{2} = M_{0}^{2} + N\Omega, \ \ 
	N = L + 2N',
\end{equation}
where $M_{0}$ is the ground-state mass, $\Omega^{-1}$
 the slope parameter of linear Regge trajectories,
 and $L$ ($N'$) the orbital-angular-momentum (radial)
 quantum number. We take here a value of $\Omega$ to be
 $1.136\ \mathrm{GeV}^{2}$ for the $n\bar{n}$ meson system,
 determined from the mass-squared distance between the $\rho(770)$
 (775.8 MeV) and $a_{2}(1320)$ (1318.3 MeV) mesons and also
 take the respective ground-state masses
 obtained in \S\ref{masses} as values of the $M_{0}(n\bar{n})$.
 For the $P^{(E)}$ sector we use a mass, 1300 MeV,
 of the $\pi(1300)$ meson,
 which was assigned to the isovector $P^{(E)}$ state
 in \S\ref{expcand}. Since we have no appropriate ground state
 as input for the $P^{(N)}$ sector, due to its Nambu-Goldstone nature,
 we use a mass, 1880 MeV, of the $\pi _{2}(1880)$ meson which is
 considered as an experimental candidate for the excited
 $1^{1}D_{2}$ isovector state.
 Then we estimate masses of the excited $1P$, $1D$ and
 $2S$ states for the respective sectors of $n\bar{n}$ mesons.
 The results are presented in Table \ref{table:5}.

 \begin{table}
 \caption{Estimated masses (in units of GeV) of the $1S$, $1P$, $2S$
 and $1D$ states for the respective sectors of states
 in the $n\bar{n}$ system. Input values are underlined.}
 \label{table:5}
 \begin{center}
 \renewcommand{\arraystretch}{1.3}
	\begin{tabular}{cccccccccc} \hline \hline
	$N$ & $L$ & $P^{(N)}$ & $S_{A}^{(N)}$ & $P^{(E)}$ & $S_{B}^{(E)}$ & $V^{(NR)}$ & $A^{(N)}$ & $V^{(ER)}$ & $B^{(E)}$ \\ \hline
	& & \multicolumn{4}{c}{$1^{1}S_{0}$} & \multicolumn{4}{c}{$1^{3}S_{1}$} \\
	0 & 0 & $0^{-+}$ & $0^{++}$ & $0^{-+}$ & $0^{+-}$ & $1^{--}$ & $1^{++}$ & $1^{--}$ & $1^{+-}$ \\ \cline{3-10}
	& & -- & \underline{0.985} & \underline{1.30} & \underline{0.790} & \underline{0.776} & \underline{1.21} & \underline{1.29} & \underline{1.23} \\ \hline
	& & & & & & \multicolumn{4}{c}{$1^{3}P_{0}$}\\
	& & & & & & $0^{++}$ &  $0^{--}$ & $0^{++}$ & $0^{-+}$\\ \cline{7-10}
	& & \multicolumn{4}{c}{$1^{1}P_{1}$} & \multicolumn{4}{c}{$1^{3}P_{1}$} \\
	1 & 1 & $1^{+-}$ & $1^{--}$ & $1^{+-}$ & $1^{-+}$ & $1^{++}$ & $1^{--}$ & $1^{++}$ & $1^{-+}$ \\ \cline{7-10}
	& & \multicolumn{4}{c}{} & \multicolumn{4}{c}{$1^{3}P_{2}$} \\
	& & & & & & $2^{++}$ &  $2^{--}$ & $2^{++}$ & $2^{-+}$ \\ \cline{3-10}
	& & 1.55 & 1.45 & 1.68 & 1.33 & 1.32 & 1.61 & 1.67 & 1.63 \\ \hline
	& & \multicolumn{4}{c}{$2^{1}S_{0}$} & \multicolumn{4}{c}{$2^{3}S_{1}$} \\
	2 & 0 & $0^{-+}$ & $0^{++}$ & $0^{-+}$ & $0^{+-}$ & $1^{--}$ & $1^{++}$ & $1^{--}$ & $1^{+-}$ \\ \cline{3-10}
	& & 1.88 & 1.80 & 1.99 & 1.70 & 1.70 & 1.93 & 1.98 & 1.95 \\ \hline
	& & & & & & \multicolumn{4}{c}{$1^{3}D_{1}$}\\
	& & & & & & $1^{--}$ &  $1^{++}$ & $1^{--}$ & $1^{+-}$\\ \cline{7-10}
	& & \multicolumn{4}{c}{$1^{1}D_{2}$} & \multicolumn{4}{c}{$1^{3}D_{2}$} \\
	2 & 2 & $2^{-+}$ & $2^{++}$ & $2^{-+}$ & $2^{+-}$ & $2^{--}$ & $2^{++}$ & $2^{--}$ & $2^{+-}$ \\ \cline{7-10}
	& & \multicolumn{4}{c}{} & \multicolumn{4}{c}{$1^{3}D_{3}$} \\
	& & & & & & $3^{--}$ & $3^{++}$ & $3^{--}$ & $3^{+-}$ \\ \cline{3-10}
	& & \underline{1.88} & 1.80 & 1.99 & 1.70 & 1.70 & 1.93 & 1.98 & 1.95 \\ \hline
	\end{tabular}
  \end{center}
\end{table}

\subsection{The $\widetilde{U}(12)_{SF} \times O(3,1)_{L}$ classification
 of observed mesons below $\sim 2$ \textrm{GeV}}
 
We seek experimental candidates for the $1S$, $1P$, $2S$ and $1D$
 states out of the known mesons\footnote{They include meson states
 which are listed under the section `Further States' in Ref.~\citen{PDG2004}.}
 listed in the Particle Data Group
 2004,\cite{PDG2004}
 in addition to the states assigned in the previous sections,
 based on their $J^{PC}$ quantum numbers, measured masses
 and decay modes. The resulting possible assignments are given
 in Table \ref{table:6} for the mesons selected in this way,
 though very tentative. From this table it is found that we have
 a number of observed states which could be classified well
 in terms of the $\widetilde{U}(12)_{SF} \times O(3,1)_{L}$ scheme.

Since the $1^{-+}$ exotic states are in the $P$-wave excitation
 of the $S_{B}^{(E)}(0^{+-};1^{1}S_{0})$ and
 $B^{(E)}(1^{+-};1^{3}S_{1})$ ground
 states, their masses are given as
\begin{equation}
	M(1^{-+};1^{1}P_{1}\ S_{B}^{(E)}) = 1.33\ \mathrm{GeV}, \ \ 
	M(1^{-+};1^{3}P_{1}\ B^{(E)}) = 1.63\ \mathrm{GeV},
\end{equation}
corresponding to the respective ground-state masses,
 $0.790\ \mathrm{GeV}$ and $1.230\ \mathrm{GeV}$.
 We see that these masses are quite consistent with
 their measured values~\cite{PDG2004} as
\begin{equation}
	M_{\pi _{1}(1400)} = 1376\pm 17\ \mathrm{MeV}, \ \ 
	M_{\pi _{1}(1600)} = 1653^{+18}_{-15}\ \mathrm{MeV}.
\end{equation}

Recently, the BES collaboration has reported the observation of
 a broad $K^{+}K^{-}$ resonance $X(1576)$ with $J^{PC}=1^{--}$
 and the pole position of $1576^{+49}_{-55}(\mathrm{stat})^{+98}
 _{-91}(\mathrm{syst}) - 
 i409^{+11}_{-12}(\mathrm{stat})^{+32}_{-67}(\mathrm{syst})$ MeV
 in the decay $J/\psi \rightarrow K^{+}K^{-}\pi^{0}$.\cite{BESX}
 Subsequently, it is shown that this resonance should be an isovector
 state.\cite{GS} 
 In the $\widetilde{U}(12)_{SF} \times O(3,1)_{L}$-classification scheme,
 taking into account its measured mass of $\sim 1.5$-$1.6$ GeV,
 we have a proper place to assign the $X(1576)$,
 that is, $(1^{--};1^{3}P_{1}\ A^{(N)})$
 in the $P$-wave excitation of the $A^{(N)}(1^{++};1^{3}S_{1})$.
 Since both the quark and antiquark of this state are the same
 chirality like $V^{(N)}(1^{--};1^{3}S_{1})$,
 the $X(1576)$ and its isoscalar partners might be expected to be seen
 in the $e^{+}e^{-}$ annihilation process, as was mentioned
 in \S \ref{vector}, though in this case the production rates
 may be considerably suppressed as compared with the $S$-wave states due to
 their $P$-wave spatial wave functions.

We have further another $1^{--}$ vector multiplet in the
 $P$-wave excitation of the $S_{A}^{(N)}(0^{++};1^{1}S_{0})$
 and vector mesons belonging to this multiplet is composed
 of a quark and an antiquark
 which have the opposite chirality each other the same as
 $S_{A}^{(N)}$. Therefore, it is expected that isoscalar and
 isovector members of this multiplet would be doubly hard to
 be produced in the $e^{+}e^{-}$ annihilation process.

\begin{table}
 \caption{Possible assignments for the known mesons
 to the $1S$, $1P$, $2S$ and $1D$ states in the
 $\widetilde{U}(12)_{SF} \times O(3,1)_{L}$-classification scheme.
 Mesons in brackets are the unknown states whose masses are estimated
 in the present work.}
 \label{table:6}
 \begin{center}
 \let\tabularsize\scriptsize
 \renewcommand{\arraystretch}{1.28}
	\begin{tabular}{cccccccc} \hline \hline
	$P^{(N)}$ & $S_{A}^{(N)}$ & $P^{(E)}$ & $S_{B}^{(E)}$ & $V^{(NR)}$ & $A^{(N)}$ & $V^{(ER)}$ & $B^{(E)}$ \\ \hline
	\multicolumn{4}{c}{$1^{1}S_{0}$} & \multicolumn{4}{c}{$1^{3}S_{1}$} \\
	$0^{-+}$ & $0^{++}$ & $0^{-+}$ & $0^{+-}$ & $1^{--}$ & $1^{++}$ & $1^{--}$ & $1^{+-}$ \\ \hline
	$\pi$ & $a_{0}(980)$ & $\pi(1300)$ & $[b_{0}(790)]$  & $\rho(770)$ & $a_{1}(1260)$ & $\rho(1250)$ & $b_{1}(1235)$ \\
	$\eta$ & $\sigma$ & $\eta(1295)$ & $[h_{0}(790)]$ & $\omega(782)$ & $f_{1}(1285)$ & $\omega(1250)$ & $h_{1}(1170)$ \\
	$\eta'(958)$ & $f_{0}(980)$ & $\eta(1475)$ & $[h_{0}'(1030)]$ & $\phi(1020)$ & $f_{1}(1420)$ & $[\phi(1540)]$ & $h_{1}(1380)$ \\
	$K$ & $[\kappa'(1135)]$ & $K(1460)$ & $\kappa$ & $K^{*}(892)$ & $K_{1}(1270)$ & $K^{*}(1410)$ & $K_{1}(1400)$ \\ \hline
	& & & & \multicolumn{4}{c}{$1^{3}P_{0}$}\\
	& & & & $0^{++}$ &  $0^{--}$ & $0^{++}$ & $0^{-+}$\\ \cline{5-8}
	& & & & & & $a_{0}(1450)$\\
	& & & & $f_{0}(1370)$ & & $f_{0}(1500)$ & $\eta(1405)$\\
	& & & & & & $f_{0}(1710)$ \\
	& & & & $K^{*}_{0}(1430)$ \\ \hline
	\multicolumn{4}{c}{$1^{1}P_{1}$} & \multicolumn{4}{c}{$1^{3}P_{1}$} \\
	$1^{+-}$ & $1^{--}$ & $1^{+-}$ & $1^{-+}$ & $1^{++}$ & $1^{--}$ & $1^{++}$ & $1^{-+}$ \\ \hline
	& $\rho(1450)$ & & $\pi_{1}(1400)$ &   & $X(1576)^{\textrm{a)}}$ & $a_{1}(1640)$ & $\pi_{1}(1600)$ \\
	$h_{1}(1595)$ & $\omega(1420)$ \\
	& $\phi(1680)$ & & & $f_{1}(1510)$ & & \\
	$K_{1}(1650)$ &   &   &   &   & $K^{*}(1680)$ \\ \hline
	& & & & \multicolumn{4}{c}{$1^{3}P_{2}$} \\
	& & & & $2^{++}$ &  $2^{--}$ & $2^{++}$ & $2^{-+}$ \\ \cline{5-8}
	& & & & $a_{2}(1320)$ & & $a_{2}(1700)$ & $\pi_{2}(1670)$ \\
	& & & & $f_{2}(1270)$  & & $f_{2}(1640)$ & $\eta_{2}(1645)$ \\
	& & & & $f '_{2}(1525)$\\
	& & & & $K^{*}_{2}(1430)$ & $K_{2}(1580)$ & & $K_{2}(1770)$ \\ \hline
	\multicolumn{4}{c}{$2^{1}S_{0}$} & \multicolumn{4}{c}{$2^{3}S_{1}$} \\
	$0^{-+}$ & $0^{++}$ & $0^{-+}$ & $0^{+-}$ & $1^{--}$ & $1^{++}$ & $1^{--}$ & $1^{+-}$ \\ \hline
	$\pi(1800)$ & & & & $\rho(1700)$ & & $\rho(1900)$ \\
	$\eta(1760)$ & $f_{0}(1790)^{\textrm{b)}}$ & & & $\omega(1650)$ \\
	& \\
	$K(1830)$ & $K^{*}_{0}(1950) $\\ \hline
	& & & & \multicolumn{4}{c}{$1^{3}D_{1}$}\\
	& & & & $1^{--}$ &  $1^{++}$ & $1^{--}$ & $1^{+-}$\\ \cline{5-8}
	& & & & & $a_{1}(1930)$ & $\rho(1965)$ & $b_{1}(1960)$\\
	& & & & & $f_{1}(1970)$ & $\omega(1960)$ & $h_{1}(1965)$\\
	& \\
	& \\ \hline
	\multicolumn{4}{c}{$1^{1}D_{2}$} & \multicolumn{4}{c}{$1^{3}D_{2}$} \\
	$2^{-+}$ & $2^{++}$ & $2^{-+}$ & $2^{+-}$ & $2^{--}$ & $2^{++}$ & $2^{--}$ & $2^{+-}$ \\ \hline
	$\pi_{2}(1880)$ & & $\pi_{2}(2005)$ &   &   & $a_{2}(1990)$ & $\rho_{2}(1940)$ \\
	$\eta_{2}(1870)$ & $f_{2}(1810)$ & $\eta_{2}(2030)$ & & & $f_{2}1910$ & $\omega_{2}(1975)$ \\
	& $f_{2}(1950)$ &   &   &  & $f_{2}(2010)$ \\
	& $K^{*}_{2}(1980)$ &   &   & $K_{2}(1820)$ \\ \hline
	\end{tabular}
  \end{center}
\end{table} 


\begin{table}[t]
\setcounter{table}{5}
\caption{(\textit{Continued}).}
 \begin{center}
 \let\tabularsize\scriptsize
 \renewcommand{\arraystretch}{1.3}
	\begin{tabular}{cccccccc} \hline \hline
	$P^{(N)}$ & $S_{A}^{(N)}$ & $P^{(E)}$ & $S_{B}^{(E)}$ & $V^{(NR)}$ & $A^{(N)}$ & $V^{(ER)}$ & $B^{(E)}$ \\ \hline
	& & & & \multicolumn{4}{c}{$1^{3}D_{3}$} \\
	& & & & $3^{--}$ &  $3^{++}$ & $3^{--}$ & $3^{+-}$ \\ \cline{5-8}
	& & & & $\rho_{3}(1690)$ & $a_{3}(1875)$ & $\rho_{3}(1990)$ & $b_{3}(2025)$ \\
	& & & & $\omega_{3}(1670)$  & & $\omega_{3}(1945)$ & $h_{3}(2025)$ \\
	& & & & $\phi_{3}(1850)$ \\
	& & & & $K^{*}_{3}(1780)$ \\ \hline
	\multicolumn{8}{l}{a) This state is taken from Ref.~\citen{BESX}.} \\
	\multicolumn{8}{l}{b) This state is taken from Ref.~\citen{BESf0}.} 
	\end{tabular}
  \end{center}
\end{table}

\section{Concluding remarks}

We have proposed the possible assignments for a number of
 observed mesons below $\sim 2$ GeV in the
 $\widetilde{U}(12)_{SF} \times O(3,1)_{L}$-classification
 scheme. Considering the phenomenological mixing scheme of
 the normal and extra states, we estimated the masses of
 missing members of the ground-state multiplets and also
 the masses of the excited $1P$, $1D$ and $2S$ states.
 We see that enigmatic states, such as the light scalar
 $\sigma$, $\kappa$, $a_{0}(980)$ and $f_{0}(980)$,
 the exotic $1^{-+}$ states $\pi _{1}(1400)$ and $\pi _{1}(1600)$,
 the low-mass vector $\rho(1250)$ and $\omega(1250)$,
 and the unexpectedly low-mass states $D_{sJ}^{*}(2317)$
 and $D_{sJ}(2460)$
 could be classified naturally as conventional $q\bar{q}$ states
 in the $\widetilde{U}(12)_{SF} \times O(3,1)_{L}$-classification
 scheme, without resort to more exotic or farfetched interpretations
 like a multiquark, molecule or low-mass hybrid.
 Since it goes without saying that only mass spectra are
 insufficient to establish their assignments, it is important
 to examine the production and decay properties, such as pionic
 and radiative transitions, of the assigned states in the
 $\widetilde{U}(12)_{SF} \times O(3,1)_{L}$-classification
 scheme.

\section*{Acknowledgements}
I am grateful to Shin Ishida, Kunio Takamatsu and other members
of the sigma group for useful discussions.

%


\begin{thebibliography}{99}
  
\bibitem{IIM2000}
S. Ishida, M. Ishida and T. Maeda, \PTP{104,2000,785}.

\bibitem{IIYMO2005}
S. Ishida, M. Ishida, K. Yamada, T. Maeda and M. Oda,
 hep-ph/0408136.
 
\bibitem{PDG2004}
Particle Data Group, S. Eidelman et al., \PLB{592,2004,1}.

\bibitem{BES}
BES Collaboration, M.Ablikim et al.,
 \PLB{633,2006,681}.

\bibitem{Maeda}
T. Maeda, K. Yamada, M. Oda and S. Ishida, these proceedings,
 \textit{Proceedings of the Seminar on Perspectives for Studies
 of Chiral Particles at BES, IHEP, Beijing, 2006}, KEK Proceedings.

\bibitem{Ishida2003}
S. Ishida, in \textit{HADRON SPECTROSCOPY},
 edited by E. Klempt et al.,
 AIP Conference Proceedings 717 (Melville, New York, 2004), 716.

\bibitem{LASS}
LASS Collaboration, D. Aston et al.,
 SLAC-PUB-5606 (1994).

\bibitem{OBELIX}
OBELIX Collaboration, A. Bertin et al.,
 \PLB{414,1997,220}.
 
\bibitem{SND}
M. N. Achasov et al.,
 \PLB{462,1999,365}; \PRD{66,2002,032001}; \PRD{68,2003,052006}.

\bibitem{BABAR}
BABAR Collaboration, B. Aubert et al.,
 \PRD{70,2004,072004}.

\bibitem{SNDop}
M. N. Achasov et al.,
 \PLB{486,2000,29}.

\bibitem{CMD2}
R. R. Akhmetshin et al.,
 \PLB{562,2003,173}.

\bibitem{Komada}
T. Komada,
 in \textit{HADRON SPECTROSCOPY}, edited by A. Reis et al.,
 AIP Conference Proceedings 814 (Melville, New York, 2006), 458;
 these proceedings, \textit{Proceedings of the Seminar on
 Perspectives for Studies
 of Chiral Particles at BES, IHEP, Beijing, 2006}, KEK Proceedings.

\bibitem{UKQCD}
UKQCD Collaboration, C. McNeile and C. Michael,
 \PRD{74,2006,014508}.

\bibitem{Glozman}
L. Ya. Glozman,
 \PLB{587,2004,69}. \\
Thomas D. Cohen and Xiangdong Ji,
 \PRD{55,1997,6870}.

\bibitem{BESX}
BES Collaboration, M.Ablikim et al.,
 hep-ex/0606047.

\bibitem{GS}
Feng-Kun Guo and Peng-Nian Shen,
 hep-ph/0606273.

\bibitem{BESf0}
BES Collaboration, M.Ablikim et al.,
 \PLB{607,2005,243}.


\end{thebibliography}
\end{document}